# Laser-microstructured ZnO/p-Si photodetector with enhanced and broadband responsivity across the UV-Vis-NIR

*Georgios Chatzigiannakis[1,2], Angelina Jaros[3], Renaud Leturcq[4], Jörgen Jungclaus[3], Tobias Voss[3], Spiros Gardelis[2,*], Maria Kandyla[1,*]*

[1]Theoretical and Physical Chemistry Institute, National Hellenic Research Foundation, 48 Vassileos Constantinou Avenue, 11635 Athens, Greece

[2]Department of Physics, National and Kapodistrian University of Athens, Panepistimiopolis Zografos, 15784 Athens, Greece

[3]Institute of Semiconductor Technology, Braunschweig University of Technology, Hans-Sommer Strasse 66, 38106 Braunschweig, Germany

[4]Materials Research and Technology Department, Luxembourg Institute of Science and Technology, 41 Rue du Brill, L-4422 Belvaux, Luxembourg

**Abstract**

We develop ZnO/p-Si photodetectors by atomic layer deposition (ALD) of ZnO thin films on laser-microstructured silicon and we investigate their electrical and optical behavior, demonstrating high sensitivity and broadband operation. Microstructured p-type silicon was obtained by ns-laser irradiation in $SF_6$ gas, which results in the formation of quasi-ordered and uniform microspikes on the silicon surface. The irradiated silicon contains sulfur impurities, which extend its absorbance to the near infrared. A thin film of ZnO was conformally deposited on the microstructured silicon substrates by ALD. Photoluminescence measurements indicate high crystalline quality of the ZnO film after annealing. Current-voltage ($I$-$V$) measurements of the ZnO/p-Si heterodiodes in dark show a non-linear behavior with unusual high current values in reverse bias. Under illumination photocurrent is observed for reverse bias, even for wavelengths below the silicon bandgap in the case of the laser-microstructured photodetectors. Higher current values are measured for the microstructured photodetectors, compared to planar ones. Photoconductivity measurements show enhanced responsivity across the UV-Vis-NIR spectral range for the laser-microstructured devices, due to their increased surface area and light absorption.

**Introduction**

Zinc oxide (ZnO) is a very promising material for optoelectronic and transparent electronic applications due to its wide bandgap (3.37 eV), which is direct and therefore efficient for optoelectronics, its large exciton binding energy (60 meV) , and high transparency (>80%) in the visible spectral range [1],[2]. Furthermore, it can be grown in a variety of nanostructures such as nanowires, nanorods, nanobelts, nanotetrapads, etc., by low-cost and low-temperature methods [2],[3]. These properties make it a good choice for optoelectronic applications, such as blue-UV photodetection or blue-UV light emission (LEDs or lasers), complementary to GaN-based devices [3]–[5]. Despite these remarkable advantages compared to other wide bandgap semiconductors, the difficulty in introducing reproducibly high-quality p-type impurities in ZnO, which intrinsically exhibits n-type conductivity due to native defects like oxygen vacancies or zinc interstitials, remains the main drawback in optoelectronic applications based on ZnO p-n homojunctions [2],[3],[6]. Therefore, ZnO optoelectronic devices rely on heterojunctions with other p-type semiconductors, with silicon being the most common choice. Indeed, ZnO/p-Si heterojunctions have been repeatedly employed for the development of solar cells [7],[8], UV-visible photodetectors [9]–[16], and LEDs [17]–[20].

Recently, there has been a vigorous effort to develop nanostructured ZnO/p-Si photodetectors with enhanced sensitivity in the UV-visible spectral range, taking advantage of the high surface-to-volume ratio of heterojunctions consisting either of ZnO nanostructures, such as nanodots, nanorods, and nanowires, on p-type silicon [11],[12],[14],[21]-[24] , or by ZnO thin films on nanostructured silicon substrates [25]–[28]. However, some fabrication techniques of nanostructured ZnO/Si heterojunctions are not scalable, hindering the adaptation of this new technology to mass production of electronic and photonic devices. Furthermore, often ZnO nanostructures on silicon require a dielectric filling layer, in order to provide structural stabilization and prevent short-circuits between top and bottom electrodes, adding an extra sensitive step in the fabrication process [20],[22],[29]. On the other hand, deposition of ZnO thin films on nanostructured silicon does not need stabilization or electric insulation between top and bottom electrodes if the ZnO film is deposited conformally, since conformal coating of the entire silicon surface with ZnO provides full electric insulation between the electrodes. To this end, atomic layer deposition (ALD) provides extraordinary conformality and large-area uniformity. Indeed, ALD-grown ZnO/Si radial nanowire photodiodes have been demonstrated, which rely on ZnO coating of silicon nanowire arrays, and showed enhanced UV-visible photoresponse [25],[26], however this promising approach remains largely unexplored and these devices do not demonstrate near-infrared (NIR) photoresponse, as expected by the bandgaps of silicon and ZnO. To date, silicon-based UV-Vis photodetectors have been well-developed [30] but sub-bandgap NIR responsivity remains a challenge for these devices, leading to the use of other materials with significantly higher costs. Recently, various approaches have shown increased sub-bandgap absorption for silicon-based photodetectors, most of which employ hyper-doped silicon with chalcogens, combined with laser-processing of the silicon surface [31]-[35].

In this work, we develop ZnO/p-Si photodetectors by ALD deposition of ZnO thin films on laser-microstructured silicon and we investigate their electrical and optical behavior, demonstrating high sensitivity and broadband operation. In particular, for silicon microstructuring we irradiated p-type silicon with nanosecond laser pulses in $SF_6$ gas. It has been shown that this method extends the absorbance of silicon to the NIR, below its energy bandgap [36]–[38]. Thus, microstructured silicon was used as a substrate to fabricate ZnO/p-Si photodetectors with enhanced responsivity due to an increased heterojunction area and also to extend the detection spectral range to the UV-Vis-NIR. No additional dielectric layer is required, as ZnO is conformally deposited on micro-Si and short-circuits between top and bottom electrodes are excluded. Indeed, laser-microstructured ZnO/p-Si devices show an enhanced spectral responsivity across the UV-Vis-NIR spectral range relative to planar ones. Laser processing of silicon is maskless and single-step, thus amenable to large-scale fabrication. This method can be easily extended to other homojunctions or heterojunctions with large surface area, for electronic and photonic applications, and maintain the possibility of integrating such devices in the microelectronics industry.

**Experimental section**

<u>Laser-microstructured silicon substrates</u>. Monocrystalline (100) silicon wafers (p-type, boron doped, carrier concentration $N_A = 8*10^{16}$ cm$^{-3}$, mobility $\mu_p = 14$ cm$^2$/Vs) were cleaned in an ultrasonic acetone and methanol bath for 15 min and placed in a vacuum chamber, which was evacuated to $10^{-3}$ mbar and filled with 0.6 bar of SF$_6$ gas (Supporting Information, Fig. S1). The wafers were irradiated by a Q-switched Nd:YAG laser system with pulse duration 5 ns and 532 nm wavelength at a repetition rate of 10 Hz. The laser beam was focused on the surface of silicon through a quartz window using a 20-cm focal length lens to a fluence of ~ 1 J/cm$^2$. The chamber was placed on a computer-controlled set of *xy* translation stages and raster scanned with respect to the laser beam, so that each spot on the silicon surface was irradiated by ~1000 pulses. The laser processing duration for each substrate was ~20 min.

<u>Development of ZnO/p-Si devices</u>. Thin ZnO films (200 nm) were deposited on microstructured silicon substrates by atomic layer deposition, forming a ZnO/Si heterojunction. ZnO deposition was performed in a Beneq TFS200 reactor at a chamber temperature of 150°C, using diethylzinc (99.99% from Sigma Aldrich) and deionized water as precursors. ZnO films deposited on glass have been characterized by Hall effect measurements. As-grown films are n type with a carrier density $n = (2.5 \pm 0.2)*10^{19}$ cm$^{-3}$ and a mobility $\mu_n = 32 \pm 5$ cm$^2$/Vs. Prior to deposition on p-Si, silicon substrates were etched in an aqueous solution of hydrofluoric acid (5% HF for 5 minutes) to remove the native SiO$_2$. For reference, heterojunctions were also formed on pristine flat silicon substrates under identical conditions. Back electrodes were deposited on the back surface of silicon by thermal evaporation of aluminum (Al, 250 nm), followed by thermal annealing (300 °C, 40 min) in pure nitrogen ambient ($N_2$) to reduce the Al-Si contact resistance via local diffusion of Al into silicon, which increases p-doping and forms an ohmic contact [39]. Top finger contacts were deposited on ZnO by thermal evaporation of Al (220 nm) via a shadow mask, in order to allow the photoexcitation of the diode, without subsequent thermal annealing. From Hall measurements of the ZnO films on glass substrates, we find that after annealing in $N_2$ atmosphere at 300°C for 40 minutes, both the carrier concentration and the mobility decrease by one order of magnitude, with $n_{annealed} = (3.0 \pm 0.5)*10^{18}$ cm$^{-3}$ and $\mu_{n,annealed} = 2.0 \pm 0.5$ cm$^2$/Vs. The observed reduction of carrier concentration and mobility is compatible with the reduction of conductance already observed for ALD-deposited ZnO after annealing in various atmospheres [40]. Moreover, it follows a similar trend as for Al-doped ZnO [40],[41]. The decrease of the carrier concentration can be explained by the annihilation of oxygen vacancies or zinc interstitials, which are the main origins for ZnO doping. Moreover, the carrier mobility is mostly limited to grain boundaries, and its reduction upon annealing is a direct consequence of the reduction of the carrier concentration.

<u>Characterization.</u> The heterojunction surface morphology and chemical composition were investigated with a field-emission scanning electron microscope (SEM), equipped with an energy-dispersive X-ray spectroscopy (EDS) detector. Room-temperature and low-temperature (4 K) photoluminescence (PL) measurements were carried out with a HeCd laser ($\lambda = 325$ nm, $P = 1$ mW). The integration time was 200 ms at 4 K and 500 ms at room temperature, except for the measurement shown in Fig. 2a, where the integration time was 2 s.

<u>Electrical and optoelectronic measurements.</u> The insets of Figs. 5a and 5b show a schematic of the set-up for *I–V* measurements on ZnO/p-Si heterojunctions. A bias voltage was applied using a *Keithley 230* voltage source and the current was monitored by a *Keithley 485* picoammeter. For illumination of the heterojunction, a white LED without any UV component, a 365-nm UV LED, and a xenon lamp combined with a 1550-nm IR bandpass filter were used as light sources. Capacitance-voltage (*C-V*) measurements of the heterojunctions were performed at a frequency of 1 MHz. For photoconductivity measurements, we employed a xenon lamp as the light source, which was monochromated through an Oriel ¼. 77200 monochromator in the 350 – 1000 nm spectral range. The responsivity was calculated by normalizing the photocurrent of the heterojunction at short-circuit conditions (0 V) to the power of the xenon lamp for each wavelength, using a calibrated, enhanced-UV, commercial silicon photodiode (Thorlabs SM1PD2A).

**Results and discussion**

<u>Structure and surface morphology</u>

A SEM micrograph of the microstructured ZnO/p-Si heterojunction, at side (45º) view, is shown in Fig. 1a. We observe that laser processing results in the formation of a quasi-ordered array of conical microspikes on the silicon surface, also known as "black silicon" [42]–[45]. The average height of the spikes is 50 ± 3 μm and the average half-height width 7 ± 1 μm. This results in an increase of the specific surface area of microstructured silicon by ~7 times compared with planar silicon of the same dimensions. The calculation is described in Supporting Information, Fig. S2. Using laser-microstructuring, we are able to process large areas of silicon with quasi-uniform micromorphology. The model and formation mechanism of such microstructures is described in detail in Refs. [43],[46] and is the result of melting, ablation, and interference effects that occur upon nanosecond laser irradiation of silicon in $SF_6$ environment. Laser processing allows the control of surface morphology by tuning the fabrication parameters, such as wavelength, pulse duration, fluence, gas or liquid environment, pressure of gas environment, and the number of incident laser pulses [37],[43],[46]. More importantly, it is a maskless and single-step process, amenable to large-scale fabrication. For the parameters employed in this work, *i.e.*, nanosecond laser-irradiation of silicon in $SF_6$ gas, we obtain only micrometric structures. Femtosecond laser-irradiation in $SF_6$ gas also results in micrometric structures albeit smaller [47], while femtosecond irradiation in liquid environments results in nanometric silicon structures [27],[42]. The presence of $SF_6$ results in the formation of sharp microspikes, as the laser pulses dissociate $SF_6$ molecules and highly reactive fluorine species etch the silicon surface while sulfur becomes incorporated into the surface (Supporting Information, Fig. S3). Other sulfur-containing gases, such as $H_2S$, also yield sharp microspikes [37].

EDS provides chemical analysis of the investigated surface. Figure 1b shows a layered SEM image with colored dots which correspond to the detected chemical elements on the surface of the microstructured heterojunction. We observe that the heterojunction consists of silicon, zinc, and oxygen. The mapping distribution of each element alone is shown in Supporting Information, Fig. S4. The distribution of zinc and oxygen indicates conformal coating of the microstructured silicon substrate with ZnO by ALD. This is crucial for the electrical operation of the ZnO/Si heterodiodes because it prevents short circuit phenomena between top and back electrodes without the use of intermediate polymeric layers, as with ZnO nanostructures on silicon substrates. SEM images of ZnO deposited on flat silicon reveal that the grains have an elongated shape with a typical length $30 - 50$ nm (Supporting Information, Fig. S5). XRD analysis of ZnO films deposited under identical conditions (same equipment, same conditions) show that they are textured polycrystalline films, with a preferential orientation of the crystal grains with the (100) crystallographic planes parallel to the surface [48].

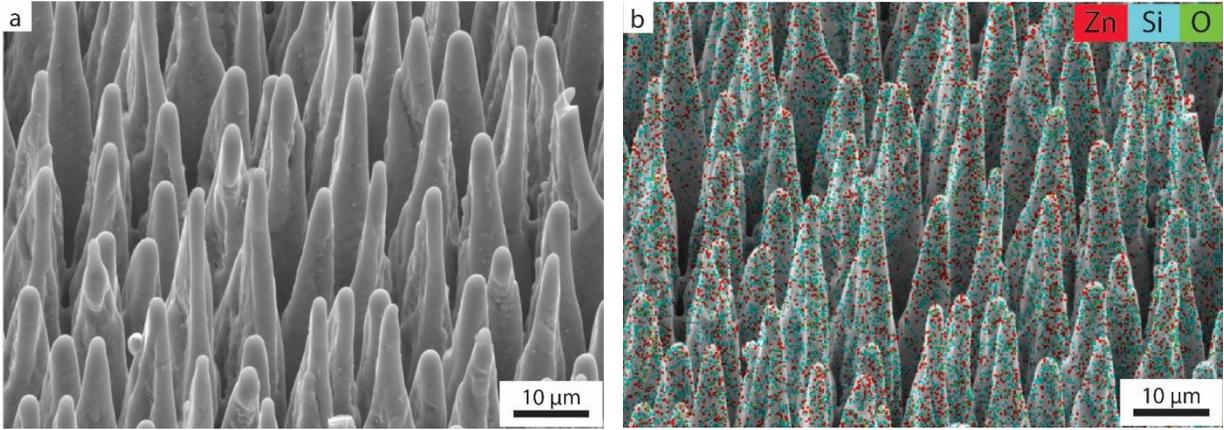

**Figure 1:** (a) SEM image of the ZnO/micro-Si surface. (b) Layered EDS image of ZnO/micro-Si. Images viewed at 45º.

Photoluminescence of ZnO on flat and micro-Si

We obtained room-temperature and low-temperature (4 K) PL spectra of as-grown ZnO films on flat silicon substrates before and after annealing. As we mention in the Experimental Section, annealing (300 ºC, $N_2$ gas ambient, 40 min) was performed after the deposition of the back Al electrode. Room-temperature PL spectra of ZnO on flat silicon before and after thermal annealing are shown in Figs. 2a and 2b. As a reference, the room-temperature PL spectrum of a ZnO single-crystal wafer (Crystec, m-plane orientation) is shown in the inset of Fig. 2b. The ZnO single crystal shows a strong UV peak at 380 nm and a weaker green band centered at 528 nm. The UV emission originates from exciton and electron-hole pair recombination processes while the green band is attributed mainly to ionized or neutral oxygen vacancies and also to other surface or bulk defects like zinc interstitials, oxygen antisites, *etc.*, [1],[2]. The as-grown ALD ZnO shows a UV peak at 389 nm and a broad visible green band centered at 492 nm (Fig. 2a). Compared to the single-crystal ZnO, the UV peak is red-shifted and broader (Table 1) while the defect-related band is also significantly broader and of higher relative intensity. On the other hand, the PL spectrum of annealed ALD ZnO (Fig. 2b) is more similar to that of single-crystal ZnO, exhibiting a strong UV peak at 384 nm and a weaker green band centered at 530 nm. Compared to the as-grown sample, the defect-related band is much weaker and narrower, indicating enhanced optical quality of ZnO since thermal annealing results in the deactivation of optical-active defects [1]. The reduction of defects after annealing is compatible with the observed reduction of carrier concentration in the Experimental Section, since free carriers are known to originate from zinc interstitials and oxygen vacancies in ZnO. Regarding the UV peak, as we can see from Table 1, it becomes narrower and blue-shifted after the ZnO film is annealed, resembling more the single-crystal UV peak, due to improved crystalline quality after annealing. Based on the UV peak of the annealed ZnO sample, and taking into account the ZnO exciton binding energy at room temperature (60 meV) [2], we estimate an optical bandgap of 3.3 eV. The reduced bandgap value, compared to the literature value of 3.37 eV, probably arises from annealing of the ZnO samples [49]. Furthermore, band-edge tailing due to the high doping density of ZnO also results in a reduced bandgap [50].

At low-temperature conditions (4 K) the annealed ALD ZnO sample shows a strong UV peak at 374 nm and a green band at 483 nm (Fig. 2c). Compared to room-temperature PL spectra, the UV peak is blue-shifted, narrower, and more intense (Table 1), while the defect-related band is also blue-shifted. The intensity increase and narrowing of the UV peak as well as the blue shift, which is due to an increased bandgap at low-temperature conditions, are expected, as reported in literature [51]. We can resolve three different sub-peaks in the main UV peak at 4 K, as shown in the inset of Fig. 2c, which can be attributed to the recombination of excitons bound to donors, longitudinal optical (LO) phonon replicas, two-electron transitions, *etc.*, [1]–[4],[51].

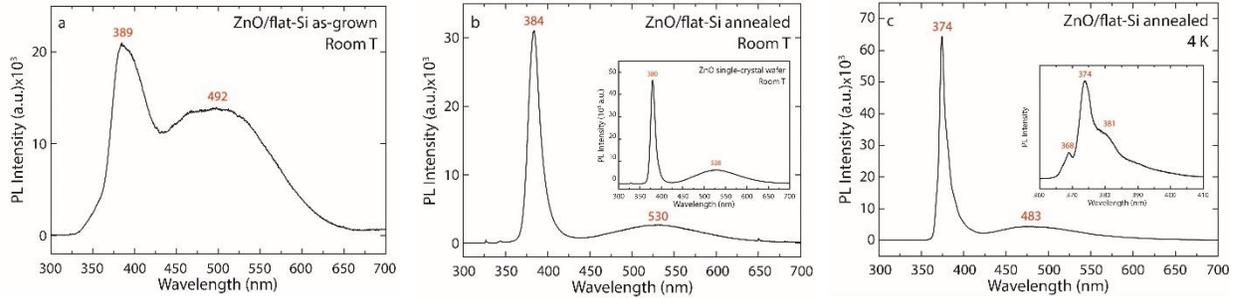

**Figure 2:** Room-temperature photoluminescence spectra of ZnO on flat silicon (a) before and (b) after thermal annealing. Inset shows the room-temperature photoluminescence spectrum of a ZnO single-crystal wafer. (c) Low-temperature photoluminescence spectrum of ZnO on flat silicon. Inset shows the UV peak in high resolution.

Figure 3 shows the PL spectra of annealed ZnO on microstructured silicon at room-temperature and low-temperature conditions (4 K). At room temperature, ZnO on micro-Si shows a strong UV peak at 384 nm and a weaker defect band centered at 518 nm (Fig. 3a), while at 4 K the UV peak is blue-shifted to 374 nm and the defect-related band to 465 nm (Fig. 3b). In the inset of Fig. 3b we observe three UV sub-peaks at 4 K, due to the mechanisms mentioned above. Comparing the PL spectra of ZnO on flat-Si and micro-Si substrates, we observe that microstructuring of the substrate does not affect the main emission properties of ZnO, since the UV peaks appear almost the same both at room temperature and at 4 K (Table 1). However, ZnO on micro-Si shows a decreased PL intensity compared to ZnO on flat-Si, since part of the emitted light is "trapped" in the microstructured surface, as a result of multiple reflections, which correlates well with the increased absorption of micro-Si [56].

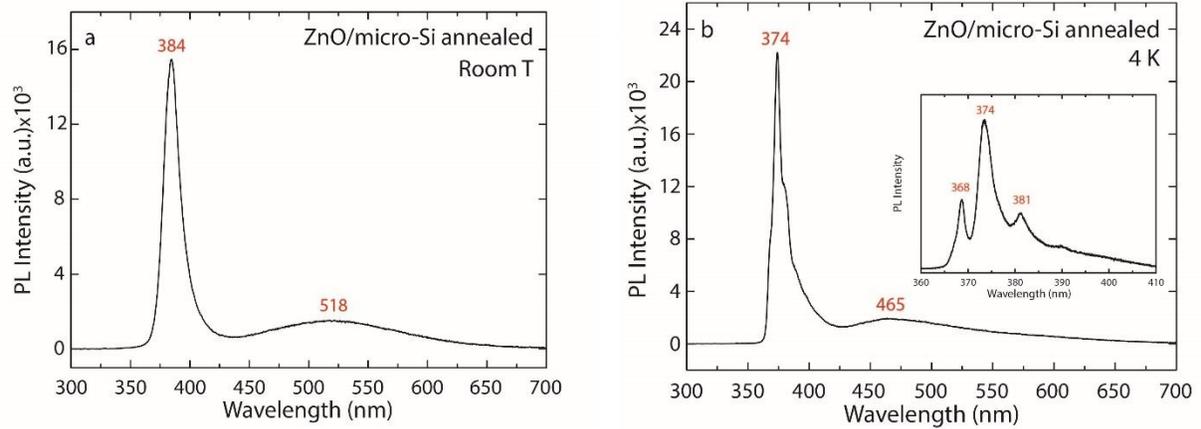

**Figure 3:** (a) Room-temperature and (b) low-temperature photoluminescence spectra of thermally annealed ZnO on microstructured silicon. Inset shows the UV peak in high resolution.

**Table 1:** Position and full width at half maximum (FWHM) of photoluminescence UV peaks.

| | Room temperature | | 4 K | |
| | UV peak position (nm) | FWHM (nm) | UV peak position (nm) | FWHM (nm) |
|---|---|---|---|---|
| **ZnO single-crystal** | 380 | 12 | 368 | 1.2 |
| **ZnO/flat-Si (as-grown)** | 389 | 35 | -- | -- |
| **ZnO/flat-Si (annealed)** | 384 | 16 | 374 | 10 |
| **ZnO/micro-Si (annealed)** | 384 | 17 | 374 | 12 |

Electrical and optoelectronic operation

First, we verify that the top Al electrode forms an ohmic contact with the ZnO layer. To this end, we deposit two Al electrodes on ZnO and perform a current-voltage (*I-V*) measurement (Fig. 4a). The *I-V* curve is linear, indicating an ohmic behavior. The ohmic character is attributed to the formation of an extremely small Schottky barrier, which enables carriers to be transferred to both sides, since the electron affinity of ZnO is comparable to the work function of Al [52],[53]. The back Al contact on p-Si substrate is also ohmic, as mentioned in the Experimental Section.

Figure 4b shows $1/C^2$ versus voltage (Mott-Schottky plot) for the ZnO/flat-Si device. We observe that $1/C^2$ decreases with decreasing reverse bias voltage and reaches a minimum value under forward bias. The linear region in the Mott-Schottky plot indicates an abrupt heterojunction between silicon and ZnO. The extraction of the linear region to the voltage axis (where $1/C^2 = 0$) indicates the built-in potential of the heterojunction, $V_{bi} = 0.57$ V, which is comparable to literature $V_{bi}$ values for ZnO/p-Si heterojunctions [8],[54],[55].

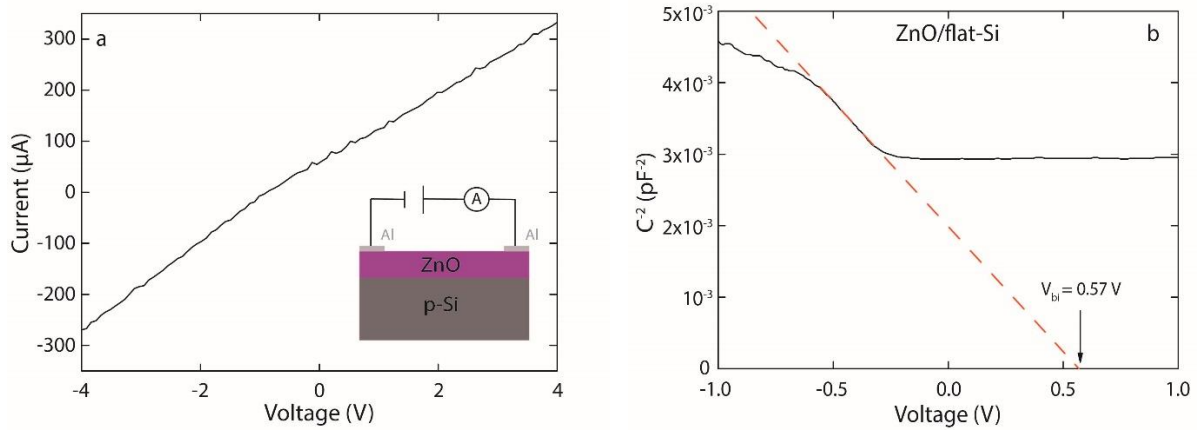

**Figure 4**: (a) *I-V* characteristic of Al electrodes on ZnO. (b) $1/C^2$ *vs.* voltage plot of the ZnO/flat-Si device (Mott-Schottky plot).

*I-V* characteristics were measured in dark and under illumination with sources covering different spectral ranges. Figure 5a shows *I-V* characteristics of the ZnO/flat-Si device in dark and under white light illumination, provided by a white-light LED with a spectral content below the energy bandgap of ZnO. The *I-V* curves are non-linear with significantly higher current values in reverse bias (negative voltage) than in forward bias (positive voltage). Under white light illumination there is a small photocurrent in reverse bias. Figure 5b shows similar *I-V* characteristics for the ZnO/micro-Si device. We also observe non-linear *I-V*s in this case with higher current values and generation of photocurrent in reverse bias. Comparing Figs. 5a and 5b, we observe that ZnO/micro-Si shows significantly higher dark and photocurrent values, compared with ZnO/flat-Si. The dark current is higher due to the increased specific surface area of the microstructured heterojunction, while the photocurrent is higher both due to increased specific surface area, which results in the generation of more photocarriers in the conduction band, and higher absorption of the microstructured surface, due to multiple reflections of the incident light. Indeed, laser-microstructured silicon alone shows higher absorption than pristine crystalline silicon [56]. Figures 5c and 5d are the semilog counterparts of Figs. 5a and 5b, respectively. At zero voltage bias we observe a higher short-circuit current ($I_{sc}$) for the microstructured device compared to the planar device. The shift of the *I-V* curves towards the forward bias region upon illumination is expected by the photovoltaic effect [8].

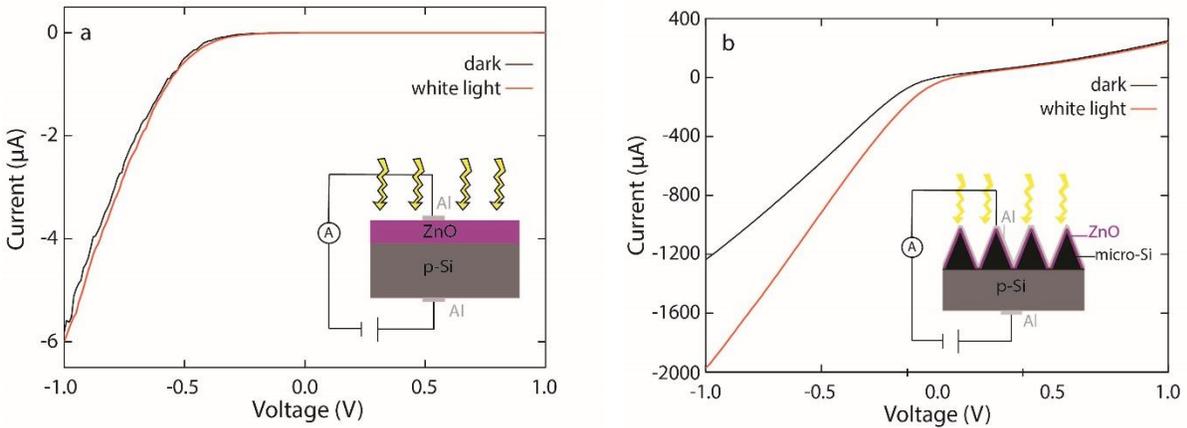

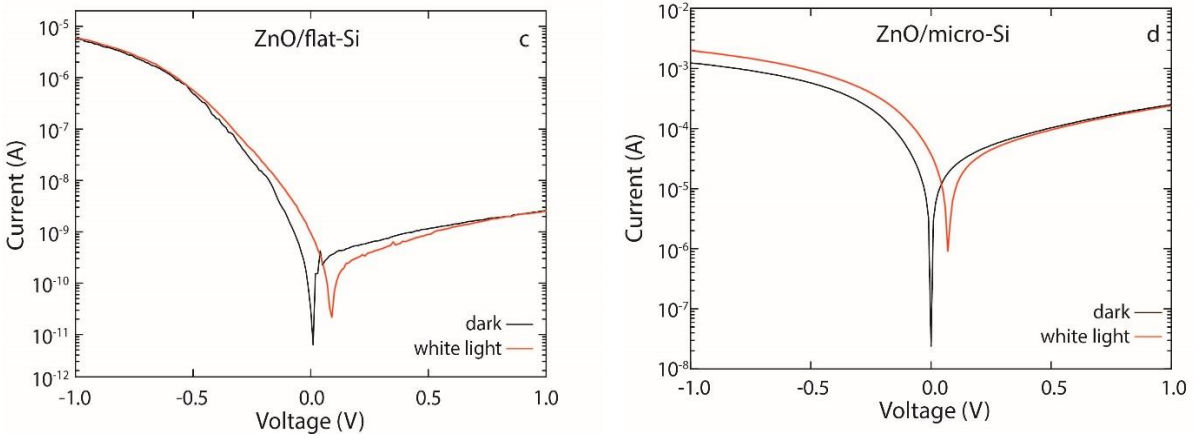

**Figure 5:** *I-V* characteristics of (a) ZnO/flat-Si and (b) ZnO/micro-Si, in dark conditions and with a white light LED source. (c),(d) Semilog plots of (a) and (b), respectively.

The increased conduction of current in the reverse rather than the forward bias for both the flat and microstructured ZnO/p-Si devices does not obey the typical rectifying diode operation. This unusual behavior cannot be attributed to Schottky junctions with the Al electrodes, since both Al/Si and Al/ZnO contacts are ohmic, as shown above. On the other hand, in this heterojunction the carrier concentration in n-ZnO is two orders of magnitude higher than in p-Si, which results in the depletion region (and band bending) extending mainly in the silicon side of the junction. Therefore, the built-in potential, $V_{bi} = 0.57$ V, that we calculated earlier is essentially the diffusion potential at the silicon surface [57]. Also, due to the bandgaps and electron affinities of the two materials, the bottom of the conduction band in ZnO lies lower in energy than that in silicon (Fig. 6). The ZnO film used in this work is a degenerate semiconductor because its Fermi level lies within the conduction band [58], since its carrier density ($3*10^{18}$ cm⁻³) is higher than the effective density of states in the ZnO conduction band ($N_c = 2*10^{18}$ cm⁻³). For the effective density of states calculation we assumed an effective electron mass of $0.19*m_o$, where $m_o$ is the free-electron mass [59]. From these considerations, it can be derived that the unusually higher current in reverse bias than in forward bias, which is reminiscent of a backward diode operation, can be due to tunneling effects from silicon to the conduction band of ZnO. Tunneling in this case is facilitated by the absence of a SiO$_2$ layer in the heterojunction, the location of Fermi level in ZnO, and a very thin ZnO depletion region. A high reverse current has been observed before in ZnO/p-Si heterojunctions with a significant carrier concentration difference between the two materials [60]. It is possible that surface defects also contribute to the reverse current, especially in the case of the microstructured device. On the contrary, under forward bias, the electron flow from ZnO to silicon is hindered by the energy barrier encountered by the electrons at the bottom of the conduction band in ZnO, the diffusion potential at the silicon surface, and by recombination and trapping at interface or bulk defects, located most probably in ZnO. Under white light illumination, more carriers are generated in the conduction band of Si due to band-to-band transitions, but also in the ZnO conduction band from energy states created in the bandgap due to bulk defects, as confirmed from PL measurements. Under reverse bias the photogenerated carriers are quickly swept away from the depletion region due to the increased internal field of the heterodiode, resulting in the generation of photocurrent [61].

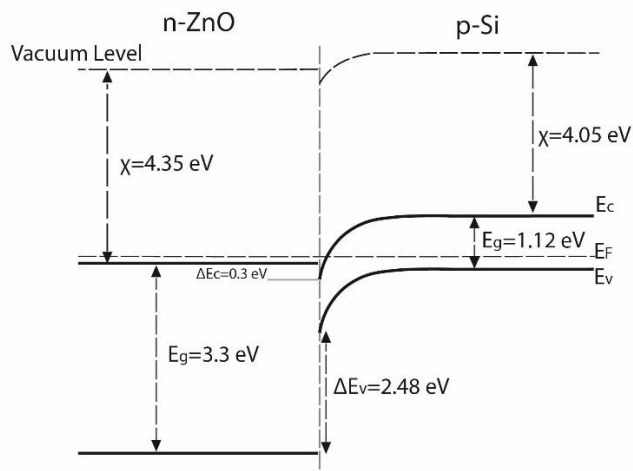

**Figure 6:** Energy band diagram of the ZnO/p-Si heterojunction in equilibrium.

Figure 7 shows the *I-V* characteristics in dark and under illumination, using a 365-nm LED source of 15.9 mW/cm$^2$ intensity, for the ZnO/flat-Si device (Fig. 7a) and the ZnO/micro-Si device (Fig. 7b). The photon energy of UV light with a wavelength of 365 nm (3.4 eV) is sufficient to excite band-to-band transitions in ZnO because it is greater than the optical bandgap of ZnO in this study (3.3 eV). Figures 7c and 7d are the semilog counterparts of Figs. 7a and 7b, respectively. We observe a higher short-circuit photocurrent for the microstructured device at zero bias and a shift of the *I-V* curves towards the forward bias region upon illumination, as for the white-light source. Comparing the $I_{sc}$ values, we observe similar short-circuit currents for the ZnO/flat-Si device either under white or UV illumination, while the ZnO/micro-Si device shows higher $I_{sc}$ under white light illumination.

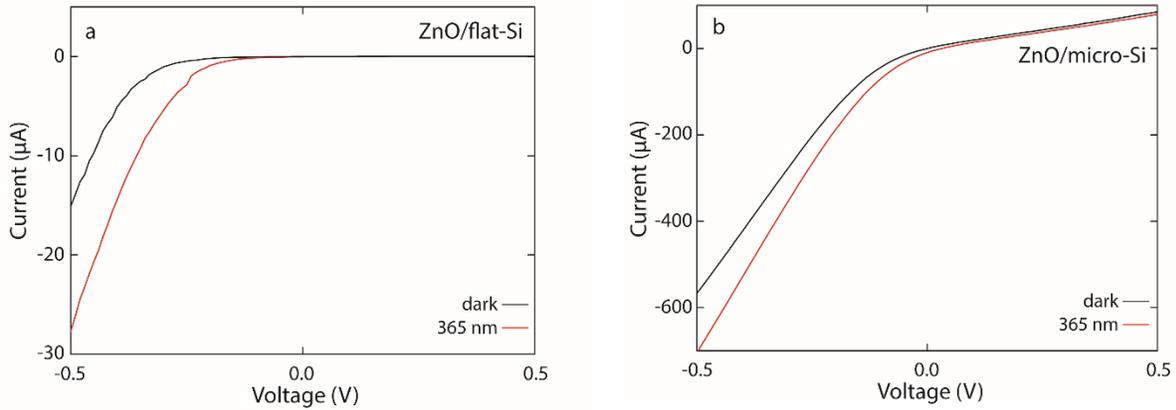

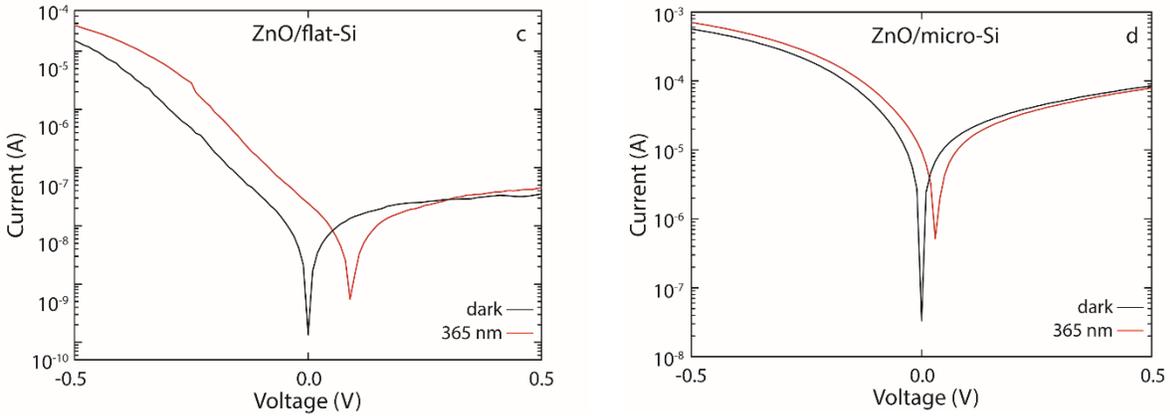

**Figure 7:** *I-V* characteristics of (a) ZnO/flat-Si and (b) ZnO/micro-Si in dark conditions and with a 365-nm LED source. (c),(d) Semilog plots of (a) and (b), respectively.

Finally, based on previous observations in the literature that laser-structured silicon in SF$_6$ absorbs near-infrared photons below the silicon optical bandgap due to sulfur states introduced into the bandgap of silicon [38],[62], we explored the possibility that the ZnO/micro-Si photodetector is sensitive to infrared illumination (1550 nm). Figure 8a shows *I-V* characteristics for the ZnO/flat-Si device in dark and under illumination, where no photocurrent is observed as both *I-V* curves coincide. This is expected as the photon energy of 1550 nm (0.8 eV) is lower than the silicon bandgap (1.1 eV), thus is not sufficient to induce band-to-band transitions in silicon. Figure 8b shows the *I-V* characteristics for ZnO/micro-Si in dark and under illumination. In contrast to the ZnO/flat-Si device, the *I-V* under 1550 nm illumination reveals the generation of photocurrent under reverse bias. Sulfur-related energy states introduced in the bandgap of silicon are optically active under IR illumination, enabling transitions of electrons to the conduction band which results in photocurrent under reverse bias. We note that below-bandgap absorbance of laser-processed silicon in SF$_6$ is known to decrease by annealing, most likely because annealing renders sulfur impurities optically inactive by causing diffusion and precipitation in grain boundaries [63].

However, it has been shown that annealing at temperatures below 300 ºC has little effect on the absorbance of this material, while annealing between 300 – 600 ºC induces a monotonic decrease in below-bandgap absorbance [38]. For this reason, we restricted the annealing temperature to 300 ºC in our devices, in order to combine ohmic-contact formation with NIR absorbance. Furthermore, it has been shown that nanosecond laser-structured samples, similar to those developed in this work, show reduction in below-bandgap absorbance by annealing to a smaller extent than femtosecond laser-structured samples, possibly because the size of the silicon microstructures is bigger in the case of nanosecond structuring, therefore NIR absorbance can be significantly amplified by multiple reflections, which are less important in femtosecond structures, the depth of which is more comparable to the NIR wavelengths [47]. Therefore, even though it is possible that annealing has affected the NIR operation of the ZnO/micro-Si photodetector, we expect this effect to be limited. HF etching for the removal of native $SiO_2$, employed here prior to ZnO deposition, has been found not to affect the morphology, sulfur concentration, or optical properties of laser-microstructured silicon in $SF_6$ [38], [64].

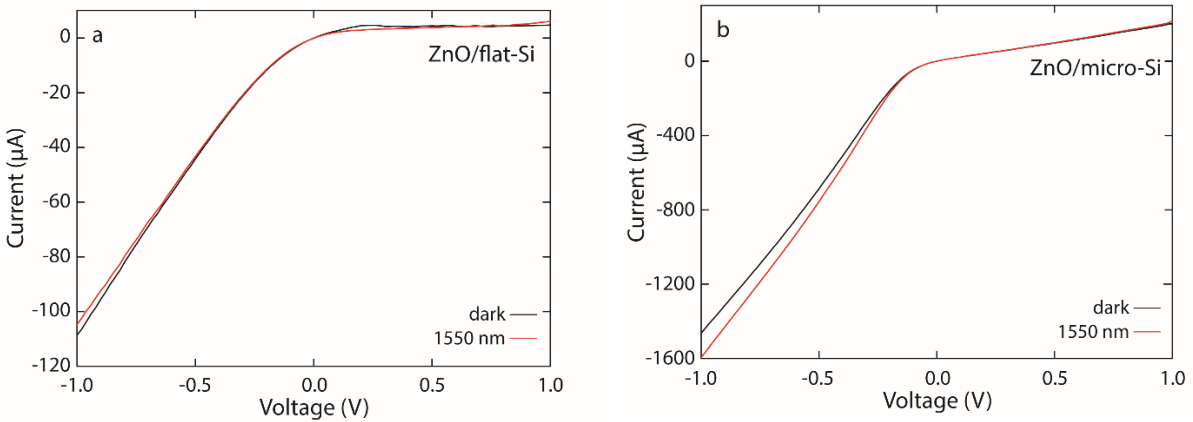

**Figure 8:** Dark and illuminated *I-V* characteristics of (a) ZnO/flat-Si and (b) ZnO/micro-Si with 1550 nm illumination.

<u>Photoconductivity</u>

We compare the responsivity of the ZnO/flat-Si and ZnO/micro-Si devices. Figure 9a shows the responsivity spectrum for the ZnO/flat-Si device at short-circuit conditions (0 V). We observe three distinct spectral regions that contribute to the photoconductivity. Region A, centered at 388 nm, is attributed to band-to-band transitions in ZnO, region B, centered at 564 nm, is attributed to above bandgap transitions in silicon but also to transitions from energy states within the bandgap of ZnO due to bulk defects, and region C, centered at 868 nm, is attributed to band edge absorption in silicon [62],[65]. We note that the centers of peaks A and B correlate well with the PL emission peaks of the annealed ZnO/flat-Si sample at room temperature, due to band-to-band and defect transitions, respectively (Fig. 2b). Figure 9b shows the responsivity spectrum of the ZnO/micro-Si device. In this case, region A (band-to-band transitions in ZnO) can be clearly distinguished whereas peaks B and C are merged into a continuum. Peak B is not clearly observed in this case, since the defect band region in ZnO/micro-Si is wider, resulting in a greater number of generated carriers in the visible region. The responsivity of ZnO/micro-Si is larger than that of ZnO/flat-Si for all wavelengths within the investigated spectral range, as shown by the ratio of the two responsivities in Fig. 9c. Therefore, the ZnO/micro-Si device exhibits enhanced responsivity across the entire UV-Vis-NIR spectrum. The responsivity enhancement is more pronounced for the NIR wavelengths where the ZnO/flat-Si device is not responsive. The enhanced responsivity of the ZnO/micro-Si device is attributed to the increased specific surface area of the heterojunction and to increased light absorption, due to multiple reflections of incident light. We also carried out photoconductivity measurements for the ZnO/micro-Si device at -0.5 V bias. We calculated the responsivity, $R$, (Supporting Information, Fig. S6) and we observe higher values with a maximum $R_{max} = 0.14$ A/W at 660 nm, compared with the maximum at 0 V, $R_{max} = 5.95$ mA/W at 890 nm.

Furthermore, there is a qualitative difference in the responsivity spectrum as the band B emerges clearly at -0.5 V in contrast to the spectrum evaluated at 0 V. In addition, we evaluated the detectivity of ZnO/micro-Si and ZnO/flat-Si devices at 0 V, using the standard thermal noise equation [59]. The detectivity values are of the order of $10^{10}$ cmHz$^{1/2}$W$^{-1}$, in agreement with simulation studies on ZnO/p-Si photodetectors [59].

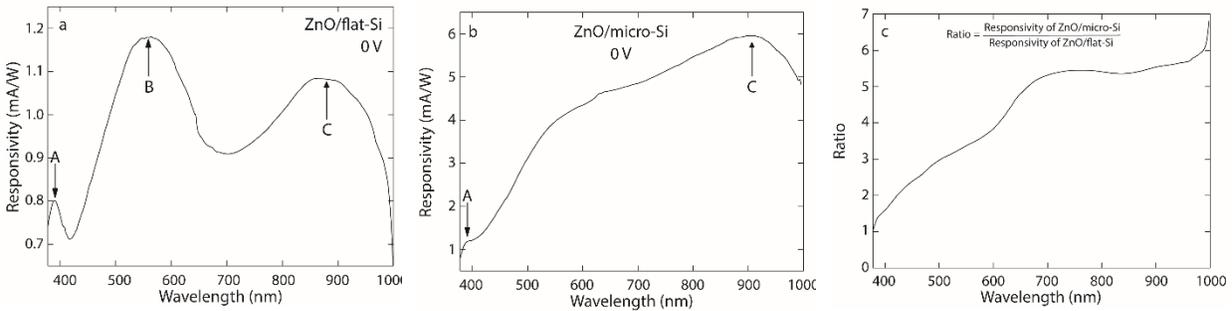

**Figure 9:** Responsivity of (a) ZnO/flat-Si and (b) ZnO/micro-Si in the UV-Vis-NIR spectral range in the absence of an external electric field. (c) Ratio of responsivities.

**Conclusions**

We developed a ZnO/micro-Si photodiode with high sensitivity and broadband response across the UV-Vis-NIR spectrum. The presence of ZnO enhances the response of the photodiode to UV radiation and the presence of laser-microstructured silicon with sulfur impurities enhances the response in the visible spectrum and enables NIR operation, contrary to most ZnO/Si or Si photodiodes. Microstructured silicon with sulfur impurities was obtained by ns-laser irradiation of p-type silicon in SF$_6$. This process is maskless and single-step, thus amenable to large-scale fabrication. Conformal coating of micro-Si with a thin ZnO film by ALD resulted in the formation of a microstructured ZnO/p-Si heterojunction, without the need of additional insulating layers for electric contacts.

We investigated the optical and electric properties, as well as the operation of the ZnO/micro-Si device as a photodetector, compared with a ZnO/flat-Si device. PL measurements reveal high optical quality and improved crystallinity of annealed ZnO on flat Si and micro-Si. The ZnO/micro-Si heterojunction shows reduced photoluminescence, as a result of multiple reflections of incident light on the microspikes, which enables enhanced light absorption, beneficial for photodetection. From the PL measurements, the bandgap of ZnO is determined as 3.3 eV. *I-V* measurements in dark reveal higher current values for the ZnO/micro-Si device, as a result of the increased junction area between silicon and ZnO. *I-V* measurements under illumination show the generation of photocurrent in the ZnO/micro-Si device in reverse bias with UV light, white light, and NIR light. Photoconductivity measurements at zero voltage reveal a significant broadband enhancement of the responsivity of the ZnO/micro-Si photodiode across the UV-Vis-NIR spectrum, even for wavelengths below the silicon bandgap, compared with the ZnO/flat-Si device. The enhancement of the UV and visible photoresponse is attributed to the increased junction area and light absorption by the microstructured device. The response of this device to NIR light is attributed to sulfur-related states incorporated in the bandgap of silicon. Such detectors can be used in a wide range of applications, requiring the detection of known sources of radiation from the ultraviolet to the infrared, including the infrared wavelengths used in telecommunications as well as visible light communications [66], without having to switch to high-cost detectors made of materials other than silicon. These results pave the way for the development of highly responsive, low-cost, broadband photodiodes compatible with monolithic integration with silicon-based electronics that go beyond the spectral capabilities of typical silicon photodetectors.

**Associated Content**

**Supporting Information**

Experimental setup for laser-microstructured silicon substrates., SEM micrograph of ZnO thin film deposited at 150°C on flat Si substrate, Calculation of the increased specific surface area, Schematic of a cone corresponding to a Si microspike. Laser-assisted $SF_6$ etching of silicon, Layered EDS image of ZnO/micro-Si. Mapping distribution of the detected elements, The responsivity spectrum of ZnO/micro-Si device under -0.5 V bias. This material is available free of charge via the Internet at http://pubs.acs.org.

**Author information**

**Corresponding Author**

*Corresponding authors: sgardelis@phys.uoa.gr; kandyla@eie.gr

**Author Contributions**

The manuscript was written through contributions of all authors. All authors have given approval to the final version of the manuscript.

**Notes**

The authors declare no conflict of interest.

**Acknowledgments**

We acknowledge support of this work by the project "Advanced Materials and Devices" (MIS 5002409) which is implemented under the "Action for the Strategic Development on the Research and Technological Sector", funded by the Operational Programme "Competitiveness, Entrepreneurship and Innovation" (NSRF 2014-2020) and co-financed by Greece and the European Union (European Regional Development Fund). We gratefully acknowledge support by the Deutsche Forschungsgemeinschaft (DFG, German Research Foundation) Research Training Group GrK1952/2 'Metrology for Complex Nanosystems'. We also thank Dr. C. Chochos for his help with SEM and EDS measurements.

**Table of Contents (TOC) Graphic**

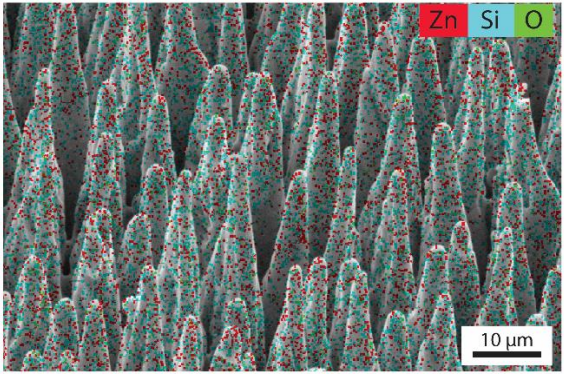
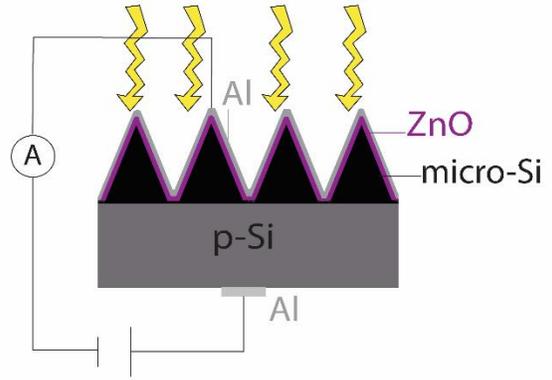